\documentclass[english,aps,manuscript]{article}
\usepackage[T1]{fontenc}
\usepackage[latin9]{inputenc}

\makeatletter

\providecommand{\tabularnewline}{\\}

\newcommand{\lyxaddress}[1]{
\par {\raggedright #1
\vspace{1.4em}
\noindent\par}
}

\usepackage{babel}
\makeatother

\begin{document}

\title{Quantum Gravity on Neutrino Mass Square difference}

\author{Bipin Singh Koranga }

\maketitle

\lyxaddress{Department of Physics, Kirori Mal college (University of Delhi,)
Delhi-110007, India}

\begin{abstract}
We consider non-renormalizable interaction term as a perturbation
of the neutrino mass matrix. We assume that the neutrino masses and
mixing arise through physics at a scale intermediate between Planck
scale and the electroweak breaking scale. We also assume that, just
above the electroweak breaking scale, neutrino masses are nearly degenerate
and their mixing is bi-maximal. Quantum gravity (Planck scale effects)
lead to an effective $SU(2)_{L}\times U(1)$ invariant dimension-5
Lagrangian involving neutrino and Higgs fields. On  symmetry breaking,
this operator gives rise to correction to the above masses and mixing.
The gravitational interaction $M_{X}=M_{pl}$, we find that for degenerate
neutrino mass spectrum, the considered perturbation term change the
$\Delta_{21}^{'}$and $\Delta_{31}^{'}$mass square difference is
unchanged above GUT scale. The nature of gravitational interaction
demands that the element of this perturbation matrix should be independent
of flavor indices. In this letter, we study the quantum gravity effects
on neutrino mass square difference, namely modified dispersion relation
for neutrino mass square differences..
\end{abstract}

\section{Introduction}

The existence of neutrino mass and mixings is experimentally well
confirmed, therefore the theortical understanding of these quantitities
is one of the most important issues for particle physics. One of most
sensitive probe of quantum gravity phenomena are neutrinos {[}1,2].
In recent year the subject of quantum gravity phenomenology has rapid
growth complementary theoretical work. Theoretical extension of the
standard model of particle physics, which also were expected to explain
the origin and the shape of small neutrino mass matrix. There is the
possible existence of non-renormalizable gravitational interaction.
Those interaction could have an influence on the neutrino sector {[}3,4,5,6,7].
In this letter we study the quantum gravity effects on neutrino mass
square diffferences. The outline of the article is as follows. In
Section 2, we briefly discuss neutrino mass square difference due
to Planck scale effects. In Section 3, we discuss about numerical
result. In Section 4, we present our conclusions.

\section{Neutrino Mass Square Difference by Perturbation Approach}

The neutrino mass matrix is assumed to be generated by the see saw
mechanism {[}9,10]. The effective gravitational interaction of neutrino
with Higgs field can be expressed as $SU(2)_{L}\times U(1)$ invariant
dimension-5 operator {[}8],

\begin{equation}
L_{grav}=\frac{\lambda_{\alpha\beta}}{M_{pl}}(\psi_{A\alpha}\epsilon\psi_{C})C_{ab}^{-1}(\psi_{B\beta}\epsilon_{BD}\psi_{D})+h.c.\end{equation}

Here and every where we use Greek indices $\alpha,\,\beta$ for the
flavor states and Latin indices i, j, k for the mass states. In the
above equation $\psi_{\alpha}=(\nu_{\alpha},l_{\alpha})$is the lepton
doublet, $\phi=(\phi^{+},\phi^{o})$is the Higgs doublet and $M_{pl}=1.2\times10^{19}GeV$
is the Planck mass $\lambda$ is a $3\times3$ matrix in a flavor
space with each elements $O(1)$. The Lorentz indices $a,b=1,2,3,4$
are contracted with the charge conjugation matrix $C$ and the $SU(2)_{L}$
isospin indices $A,B,C,D=1,2$ are contracted with $\epsilon=i\sigma_{2},\,\,\sigma_{m}(m=1,2,3)$are
the Pauli matrices. After spontaneous electroweak symmetry breaking
the Lagrangian in eq(1) generated additional term of neutrino mass
matrix

\begin{equation}
L_{mass}=\frac{v^{2}}{M_{pl}}\lambda_{\alpha\beta}\nu_{\alpha}C^{-1}\nu_{\beta},\end{equation}

where $v=174GeV$ is the $VEV$ of electroweak symmetric breaking.
We assume that the gravitational interaction is''flavor blind''
that is $\lambda_{\alpha\beta}$ is independent of $\alpha,\,\beta\,$indices.
Thus the Planck scale contribution to the neutrino mass matrix is

\begin{equation}
\mu\lambda=\mu\left(\begin{array}{ccc}
1 & 1 & 1\\
1 & 1 & 1\\
1 & 1 & 1\end{array}\right),\end{equation}

where the scale $\mu$ is 

\begin{equation}
\mu=\frac{v^{2}}{M_{pl}}=2.5\times10^{-6}eV.\end{equation}

We take eq(3) as perturbation to the main part of the neutrino mass
matrix, that is generated by GUT dynamics. We treat M as the unperturbed
($0^{th}$ order) mass matrix in the mass eigenbasis. Let U be the
mixing matrix at $0^{th}$ order. Then the corresponding $0^{th}$
order mass matrix) $M$~in flavour space {[}4] given by

\begin{equation}
\mathbf{M}=U^{*}diag(M_{i})U^{\dagger},\end{equation}

where, $U_{\alpha i}$ is the usual mixing matrix and $M_{i}$ , the
neutrino masses is generated by Grand unified theory. Most of the
parameter related to neutrino oscillation are known, the major expectation
is given by the mixing elements $U_{e3}.$ We adopt the usual parametrization.

\begin{equation}
\frac{|U_{e2}|}{|U_{e1}|}=tan\theta_{12},\end{equation}

\begin{equation}
\frac{|U_{\mu3}|}{|U_{\tau3}|}=tan\theta_{23},\end{equation}

\begin{equation}
|U_{e3}|=sin\theta_{13}.\end{equation}

In term of the above mixing angles, the mixing matrix is

\begin{equation}
U=diag(e^{if1},e^{if2},e^{if3})R(\theta_{23})\Delta R(\theta_{13})\Delta^{*}R(\theta_{12})diag(e^{ia1},e^{ia2},1).\end{equation}

The matrix $\Delta=diag(e^{\frac{1\delta}{2}},1,e^{\frac{-i\delta}{2}}$)
contains the Dirac phase. This leads to CP violation in neutrino oscillation
$a1$ and $a2$ are the so called Majoring phase, which effects the
neutrino less double beta decay. $f1,$ $f2$ and $f3$ are usually
absorbed as a part of the definition of the charge lepton field. Due
to Planck scale effects on neutrino mixing the new mixing matrix can
be written as {[}4]

\[
U^{'}=U(1+i\delta\theta),\]

\[
\left(\begin{array}{ccc}
U_{e1} & U_{e2} & U_{e3}\\
U_{\mu1} & U_{\mu2} & U_{\mu3}\\
U_{\tau1} & U_{\tau2} & U_{\tau3}\end{array}\right)\]

\begin{equation}
+i\left(\begin{array}{ccc}
U_{e2}\delta\theta_{12}^{*}+U_{e3}\delta\theta_{23,}^{*} & U_{e1}\delta\theta_{12}+U_{e3}\delta\theta_{23}^{*}, & U_{e1}\delta\theta_{13}+U_{e3}\delta\theta_{23}^{*}\\
U_{\mu2}\delta\theta_{12}^{*}+U_{\mu3}\delta\theta_{23,}^{*} & U_{\mu1}\delta\theta_{12}+U_{\mu3}\delta\theta_{23}^{*}, & U_{\mu1}\delta\theta_{13}+U_{\mu3}\delta\theta_{23}^{*}\\
U_{\tau2}\delta\theta_{12}^{*}+U_{\tau3}\delta\theta_{23}^{*}, & U_{\tau1}\delta\theta_{12}+U_{\tau3}\delta\theta_{23}^{*}, & U_{\tau1}\delta\theta_{13}+U_{\tau3}\delta\theta_{23}^{*}\end{array}\right).\end{equation}

Where $\delta\theta$ is a hermition matrix that is first order in
$\mu${[}11]. The first order mass square difference $\Delta M_{ij}^{2}=M_{i}^{2}-M_{j}^{2},$get
modified {[}12] as

\begin{equation}
\Delta_{ij}^{'}=\Delta_{ij}+2(M_{i}Re(m_{ii})-M_{j}Re(m_{jj}),\end{equation}

where

\[
m=\mu U^{t}\lambda U,\]

\[
\mu=\frac{v^{2}}{M_{pl}}=2.5\times10^{-6}eV.\]

The change in the elements of the mixing matrix, which we parametrized
by $\delta\theta${[}16], is given by

\begin{equation}
\delta\theta_{ij}=\frac{iRe(m_{jj})(M_{i}+M_{j})-Im(m_{jj})(M_{i}-M_{j})}{\Delta M_{ij}^{'^{2}}}.\end{equation}

The above equation determine only the off diagonal elements of matrix
$\delta\theta_{ij}$. The diagonal element of $\delta\theta_{ij}$
can be set to zero by phase invariance. 

Using Eq(10), we can calculate neutrino mixing angle due to Planck
scale effects,

\begin{equation}
\frac{|U_{e2}^{'}|}{|U_{e1}^{'}|}=tan\theta_{12}^{'},\end{equation}

\begin{equation}
\frac{|U_{\mu3}^{'}|}{|U_{\tau3}^{'}|}=tan\theta_{23}^{'},\end{equation}

\begin{equation}
|U_{e3}^{'}|=sin\theta._{13}^{'}\end{equation}

For degenerate neutrinos, $M_{3}-M_{1}\cong M_{3}-M_{2}\gg M_{2}-M_{1},$
because $\Delta_{31}\cong\Delta_{32}\gg\Delta_{21}.$

\section{Numerical Results}

Note from eq(11) that the correction term depends on cruically the
type of neutrino mass spectrum. For a hierarchical or inverse hierarchical
spectrum the correction is negligible. Hence, we consider a degenerated
neutrino Mass spectrum and take the common neutrino mass to be 2 eV,
which is upper limit of tritium beta decay spectrum {[}13]. From definition
of the matrix $m$ in eq(11), we find

\[
m_{11}=\mu e^{i2a_{1}}(U_{e1}e^{if1}+U_{\mu1}e^{if2}+U_{\tau1}e^{if3})^{2}\]

\[
m_{22}=\mu e^{i2a_{2}}(U_{e2}e^{if1}+U_{\mu2}e^{if2}+U_{\tau2}e^{if3})^{2}\]

The contribution of the term in the Planck scale correction, $\epsilon=2(M_{i}Re(m_{11})-M_{j}Re(m_{22}),$
can be additive or subtractive depending on the values of the phase
$a_{1},$ $a_{2}$ and phase $f_{i}$. In our calculation, we used
mixing angle as $\theta_{12}=34^{o},\,\theta_{23}=45^{o}$ ,$\theta_{13}=10^{o}.$
and $\delta=0^{o}.$We have taken $\Delta_{31}=0.002eV^{2}${[}14]
and $\Delta_{21}=0.00008eV^{2}${[}15]. For simplicity, we have set
the charge lepton phases $f_{1}=f_{2}=f_{3}=0.$ Since we have set
$\theta_{13}=0,$the Dirac phase $\delta$ drops out of the $0^{th}$order
mixing matrix. In table (1.0) we list the modified mass square difference
terms for some sample value of $a_{1}$ and $a_{2}$.

\begin{table}
\begin{tabular}{|c|c|c|c|}
\hline 
$a_{1}$ & $a_{2}$ & $\Delta_{21}^{'}$ & $\Delta_{31}^{'}$\tabularnewline
\hline
\hline 
$0^{o}$ & $0^{o}$ & $7.6\times10^{-5}eV^{2}$ & $2.0\times10^{-3}eV^{2}$\tabularnewline
\hline 
$0^{o}$ & $45^{o}$ & $7.3\times10^{-5}eV^{2}$ & $2.0\times10^{-3}eV^{2}$\tabularnewline
\hline 
$0^{o}$ & $90^{o}$ & $6.9\times10^{-5}eV^{2}$ & $2.0\times10^{-3}eV^{2}$\tabularnewline
\hline 
$0^{o}$ & $145^{o}$ & $7.3\times10^{-5}eV^{2}$ & $2.0\times10^{-3}eV^{2}$\tabularnewline
\hline 
$0^{o}$ & $180^{o}$ & $7.6\times10^{-5}eV^{2}$ & $2.0\times10^{-3}eV^{2}$\tabularnewline
\hline 
$45^{o}$ & $0^{o}$ & $8.3\times10^{-5}eV^{2}$ & $2.1\times10^{-3}eV^{2}$\tabularnewline
\hline 
$45^{o}$ & $45^{o}$ & $8.3\times10^{-5}eV^{2}$ & $2.1\times10^{-3}eV^{2}$\tabularnewline
\hline 
$45^{o}$ & $90^{o}$ & $7.6\times10^{-5}eV^{2}$ & $2.1\times10^{-3}eV^{2}$\tabularnewline
\hline 
$45^{o}$ & $135^{o}$ & $7.9\times10^{-5}eV^{2}$ & $2.1\times10^{-3}eV^{2}$\tabularnewline
\hline 
$45^{o}$ & $180^{o}$ & $8.3\times10^{-5}eV^{2}$ & $2.1\times10^{-3}eV^{2}$\tabularnewline
\hline 
$90^{o}$ & $0^{o}$ & $9.0\times10^{-5}eV^{2}$ & $2.1\times10^{-3}eV^{2}$\tabularnewline
\hline 
$90^{o}$ & $45^{o}$ & $8.6\times10^{-5}eV^{2}$ & $2.1\times10^{-3}eV^{2}$\tabularnewline
\hline 
$90^{o}$ & $90^{o}$ & $8.3\times10^{-5}eV^{2}$ & $2.1\times10^{-3}eV^{2}$\tabularnewline
\hline 
$90^{o}$ & $135^{o}$ & $8.6\times10^{-5}eV^{2}$ & $2.1\times10^{-3}eV^{2}$\tabularnewline
\hline 
$90^{o}$ & $180^{o}$ & $9.0\times10^{-5}eV^{2}$ & $2.1\times10^{-3}eV^{2}$\tabularnewline
\hline 
$135^{o}$ & $0^{o}$ & $8.3\times10^{-5}eV^{2}$ & $2.1\times10^{-3}eV^{2}$\tabularnewline
\hline 
$135^{o}$ & $45^{o}$ & $7.9\times10^{-5}eV^{2}$ & $2.1\times10^{-3}eV^{2}$\tabularnewline
\hline 
$135^{o}$ & $90^{o}$ & $7.6\times10^{-5}eV^{2}$ & $2.1\times10^{-3}eV^{2}$\tabularnewline
\hline 
$135^{o}$ & $135^{o}$ & $7.9\times10^{-5}eV^{2}$ & $2.1\times10^{-3}eV^{2}$\tabularnewline
\hline 
$135^{o}$ & $180^{o}$ & $8.3\times10^{-5}eV^{2}$ & $2.1\times10^{-3}eV^{2}$\tabularnewline
\hline 
$180^{o}$ & $0^{o}$ & $7.6\times10^{-5}eV^{2}$ & $2.0\times10^{-3}eV^{2}$\tabularnewline
\hline 
$180^{o}$ & $45^{o}$ & $7.3\times10^{-5}eV^{2}$ & $2.0\times10^{-3}eV^{2}$\tabularnewline
\hline 
$180^{o}$ & $90^{o}$ & $6.9\times10^{-5}eV^{2}$ & $2.0\times10^{-3}eV^{2}$\tabularnewline
\hline 
$180^{o}$ & $135^{o}$ & $7.3\times10^{-5}eV^{2}$ & $2.0\times10^{-3}eV^{2}$\tabularnewline
\hline 
$180^{o}$ & $180^{o}$ & $7.6\times10^{-5}eV^{2}$ & $2.0\times10^{-3}eV^{2}$\tabularnewline
\hline
\end{tabular}

\caption{The modified mass square difference term for various value of phase.
Input value are $\Delta_{31}=2.0\times10^{-3}eV^{2}$, $\Delta_{21}=8.0\times10^{-8}eV^{2}$,
$\theta_{12}=34^{o},\,\theta_{23}=45^{o}$ ,$\theta_{13}=0^{o}$}

\end{table}

\section{Conclusions}

It is expected that a higher scale generation the neutrino mass matrix,
which will eventually produce the presently observed masses and mixings.
In an attractive scenario,the mixing pattern generated by higher scale
dynamics is predicted to be bimaximal. We consider that the main part
of neutrino masses and mixing from GUT scale operator. The gravitational
interaction of lepton field with S.M Higgs field give rise to a $SU(2)_{L}\times U(1)$
invariant dimension-5 effective Lagrangian give originally by Weinberg
{[}8]. On electroweak symmetry breaking this operators leads to additional
mass terms. We considered these to be perturbation of GUT scale mass
terms. This model predict a value for the modified mass square difference
$\Delta_{21}^{'}$=$\Delta_{21}\pm(1.0+0.5)\times10^{-5}eV^{2},$which
is correspondence to Planck scale $M_{pl}\approx2.0\times10^{19}GeV.$
In addition, the solar mixing angle is predicted {[}12]. In this letter,we
studied how physics from planck scale effects the neutrino mass square
difference. We compute the modified neutrino mass square difference
due to the additional mass terms for the case of bimaximal mixing.
The change in $\Delta_{31},$due to these planck scale correction
are negligible. But the change in $\Delta_{21}^{'}$ is enough that
final value falls within the expermentally accepted region. This occurs,of
course for degenerate neutrino mass with a common mass of about 2eV.

\end{document}